%% file: iclr2020_conference.tex
\documentclass{article} % For LaTeX2e

\usepackage{graphicx}

\usepackage[square,sort&compress,comma,numbers]{natbib}
\usepackage{iclr2020_conference,times}

\input{math_commands.tex}

\usepackage{hyperref}
\usepackage{url}

\usepackage{hyperref}
\usepackage{url}
\usepackage{xcolor}
\usepackage{multirow}
\usepackage{booktabs} 

\usepackage{pifont}

\usepackage{caption}
\usepackage{subcaption}

\title{Addressing Ancestry Disparities in Genomic Medicine: A Geographic-aware Algorithm}

\author{Daniel Mas Montserrat\thanks{Work conducted during an internship at Stanford University.} \\
Purdue University\\
\And
Arvind Kumar \\
Stanford University\\
\And
Carlos Bustamante \\
Stanford University\\
\And
Alexander Ioannidis \\
Stanford University\\
}

\iclrfinalcopy % Uncomment for camera-ready version, but NOT for submission.
\begin{document}

\maketitle

\begin{abstract}
\vspace{-0.1cm}
With declining sequencing costs a promising and affordable tool is emerging in cancer diagnostics: genomics \cite{Schwarze:2020dr}. By using association studies, genomic variants that predispose patients to specific cancers can be identified, while by using tumor genomics cancer types can be characterized for targeted treatment. However, a severe disparity is rapidly emerging in this new area of precision cancer diagnosis and treatment planning, one which separates a few genetically well-characterized populations (predominantly European) from all other global populations. Here we discuss the problem of population-specific genetic associations, which is driving this disparity, and present a novel solution--coordinate-based local ancestry--for helping to address it. We demonstrate our boosting-based method on whole genome data from divergent groups across Africa and in the process observe signals that may stem from the transcontinental Bantu-expansion.

\end{abstract}
\vspace{-0.2cm}

\section{Introduction}
\vspace{-0.1cm}

Cancer genomics depends upon the identification of variants that are associated with particular types of cancers. Because such variants are deleterious, they are not typically part of the ancient standing variation spread across all humans; instead they are more recent mutations specific to particular populations. Indeed, such variants are often present prominently only in particular ethnic groups due to genetic drift \cite{Foulkes:2002cc}. In addition, most associations are mapped not to causal variants, but to more common neighboring variants that are present on genotyping arrays. Since these neighboring variants are linked to the causal variant via correlation structures (linkage) that are specific to each population, the ancestry of the genomic segment in which the correlated variant is found becomes crucial. Indeed, as a result of linkage and epistatic effects, genomic variants that are associated with cancer in one ancestry maybe have no association \cite{Wang:2018js}, or may even have an opposite association \cite{Rajabli:2018cb}, in another ancestry. This phenomenon persists even in admixed individuals possessing multiple ancestries, such as African Americans; in such individuals the ancestry (European or African) of the specific genomic fragment containing the associated variant has been found to reverse the association \cite{shortRajabli:2018cb}. This phenomenon dubbed "flip-flop," is not an unusual case, rather ancestry-specific effects in genetic association studies are the rule. For this reason, polygenic-risk scores (PRS), increasingly important to genomic cancer prediction \cite{Mavaddat:2019ix}, have been found to be several times less accurate when used on populations of different ancestry from the one on which they were trained \cite{shortMartin:2019bm}.

As a result of these ancestry specific effects, accurately identifying the ancestry of each segment of the genome is becoming increasingly crucial for genomic medicine. Such algorithms, known as local ancestry inference, have been developed both for historical population genetics \cite{Tang2006, Sundquist2008, ShortPrice:2009bga, sankararaman2008estimating, Durand:2014hj, maples2013rfmix, vaegan2019, lainet2020} and for recreational consumer ancestry products \cite{Durand:2015jx}, but none have been developed to date for the particular demands of clinical genomic medicine. Such an algorithm would need to provide ancestry not as a culturally defined label, but as continuous genetic coordinates that could be used as a covariate in predication and association algorithms. 
This method is also important for deconvolving ancestry effects in genetic association studies. To date, most genome-wide association studies (GWAS) are conducted in populations of single ancestry (typically European) to avoid confounding effects of ancestry on reversing associations. Researchers often avoid admixed populations, for instance African Americans or Hispanics, who encompass more than one ancestry, and avoid populations with too much genetic variation or too many diverse sub-populations, as is common within Africa. This has resulted in over 80\% of the individuals in GWAS studies to date stemming from European ancestry (and only 2\% from African ancestry) \cite{Sirugo:2019ie, Popejoy:2016di}. A reliable coordinate-based local ancestry algorithm would allow such studies to embrace diversity, rather than intentionally eschewing it, by allowing an additional covariate along the genome to be used (ancestry) to remove the confounding effects of ancestry-dependent genomic associations. With such a tool, medical researchers would no longer need to avoid admixed and globally diverse genetic study cohorts.

\section{Ancestry Inference}
\vspace{-0.1cm}
%\textcolor{red}{maybe we should summarize this part and explain the new projection coordinate space}

Here were present an accurate coordinate-based local ancestry inference algorithm, XGMix, that can be used for addressing ancestry-specific associations and predictions. XGMix uses modern single ancestry reference populations to accurately predict the latitude and longitude of the closest modern source population for each segment of an individual's genome. These coordinate annotations along the genome can then be used as covariates for genome-wide association studies (GWAS) and for polygenic risk score (PRS) predictions.

Estimation of an individual's ancestry, both globally and locally (i.e. assigning an ancestry estimate to each region of the chromosomal sequence), has been tackled with a wide range of methods and technologies \cite{Tang2006, Sundquist2008, ShortPrice:2009bga, sankararaman2008estimating, Durand:2014hj, maples2013rfmix, vaegan2019, lainet2020}. Local ancestry inference has traditionally been framed as a classification problem using pre-defined ancestries. Classification approaches provide discrete ancestry labels but can be highly inaccurate for neighboring populations (or population gradients) and intractable for genetically diverse populations with multiple sources. Geographical regression along the genome, although a much more challenging problem, could provide a continuous representation of ancestry capable of capturing the complexities of worldwide populations.

XGMix consists of two layers of stacked gradient boosted trees (a genomic window-specific layer and a window aggregating smoother) and can infer local-ancestry with both classification probabilities and geographical coordinates along each phased chromosome. Here we demonstrate XGMix by training on whole genomes from real individuals from the five African populations included in the 1000 genomes project \cite{10002015global}. We simulate admixed individuals of various generations using Wright-Fisher simulation \cite{maples2013rfmix} to create ground truth labels of ancestry along the genome and split this data for training and testing. As these reference African populations lie close to a single arc along the globe we estimate along this arc, getting geographic assignments for each genomic segment.

\begin{figure}[htp]
    \centering

        \includegraphics[width=0.85\textwidth]{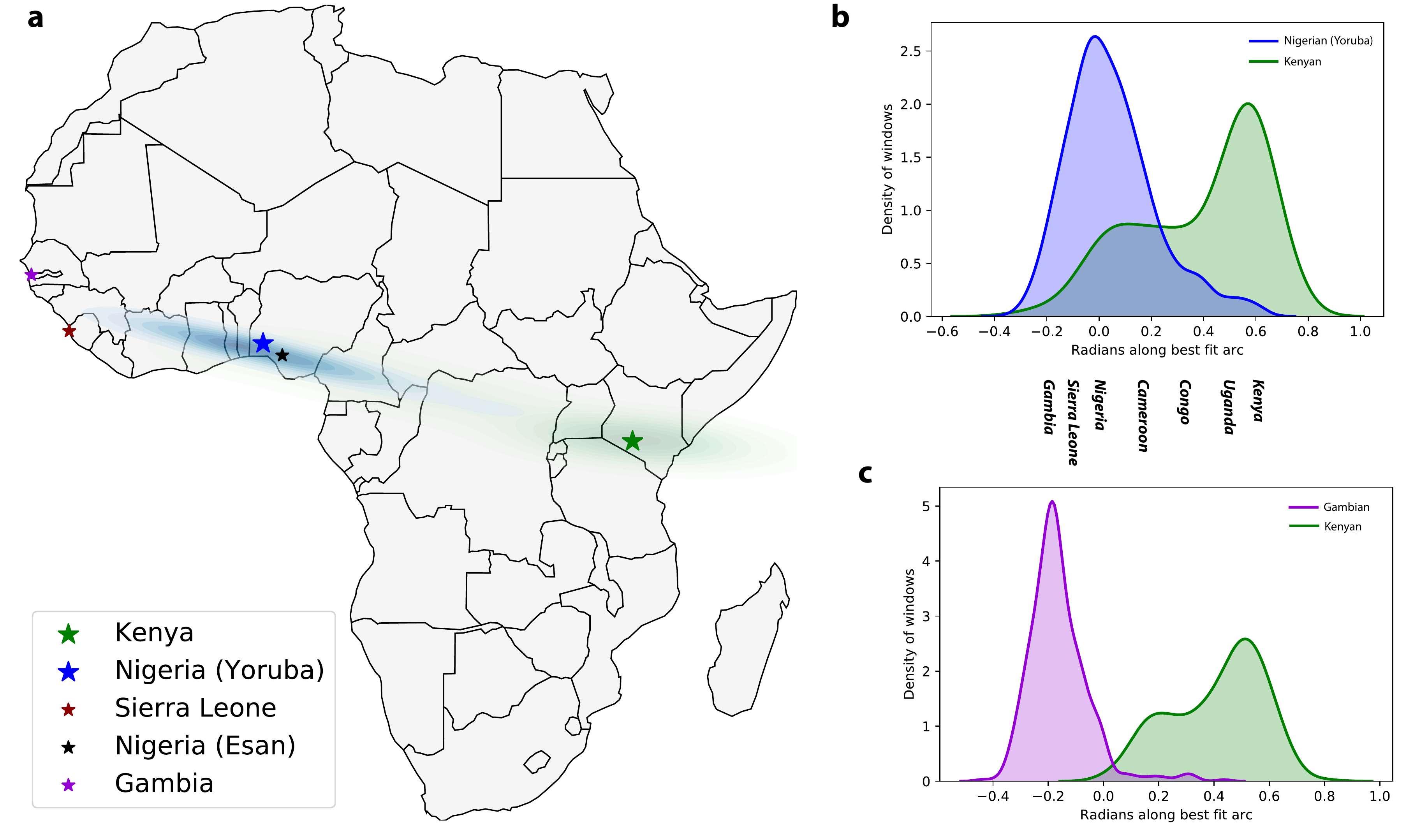}

    \caption{(a) The inferred coordinates for each genomic segment of an admixed Kenyan-Nigerian individual. The model was trained on all indicated African reference populations. (b-c) The inferred location of each genomic segment of a Kenyan-Nigerian (b) and Kenyan-Gambian (c) individual using the principal coordinate arc of the reference populations' locations. The bimodal distribution of Kenyan segments (green) may reflect the historical Bantu expansion from Cameroon into Kenya.}
    \label{fig:map}
        
\end{figure}

\bibliography{mybiblio}
\bibliographystyle{IEEEbib}

% \appendix
% \section{Appendix}

\end{document}

%% file: math_commands.tex
\usepackage{amsmath,amsfonts,bm}

\def\eqref#1{equation~\ref{#1}}
\def\1{\bm{1}}

\DeclareMathAlphabet{\mathsfit}{\encodingdefault}{\sfdefault}{m}{sl}
\SetMathAlphabet{\mathsfit}{bold}{\encodingdefault}{\sfdefault}{bx}{n}